\documentclass[12pt]{article}
\usepackage{graphics}

\begin{document}
\renewcommand \theequation{\thesection.\arabic{equation}}

\title{Deviation from standard QED at large distances:
influence of transverse dimensions of colliding beams on
bremsstrahlung}
\author{V. N. Baier
and V. M. Katkov\\
Budker Institute of Nuclear Physics,\\ Novosibirsk, 630090,
Russia}

\maketitle

\begin{abstract}
The radiation at collision of high-energy particles is formed
over a rather long distances and therefore is sensitive to an
environment. In particular the smallness of the transverse
dimensions of the colliding beams leads to suppression of
bremsstrahlung cross section for soft photons. This beam-size
effect was discovered and investigated at INP, Novosibirsk around
1980. At that time an incomplete expression for the bremsstrahlung
spectrum was calculated and used because a subtraction associated
with the extraction of pure fluctuation process was not
performed. Here this procedure is done. The complete expression
for the spectral-angular distribution of incoherent
bremsstrahlung probability is obtained. The case of Gaussian
colliding beams is investigated in details. In the case of flat
beams the expressions for the bremsstrahlung spectrum are
simplified essentially. Comparison of theory with VEPP4 and HERA
data is performed. Possible application of the effect to linear
$e^+e^-$ collider tuning is discussed.

\end{abstract}

\newpage
\section{Introduction}

The formation of the bremsstrahlung process of high-energy
particles occurs with extremely small momentum transfers. In the
space-time picture this means that the process takes place at the
large (macroscopic) distances. The longitudinal length (with
respect to the direction of the initial momentum) of formation of
the radiation usually is called the coherence (formation) length
$l_f$. For emission of a photon with energy $\omega$ the coherence
length is $l_f(\omega) \sim
\varepsilon(\varepsilon-\omega)/m^2\omega$, where $\varepsilon$
and $m$ is the energy and mass of the emitting particle ( here the
system $\hbar=c=1$ is used). If the particle experiences some
action in this length, the radiation pattern changes (in the case
when the action is the multiple scattering of the emitting
particle one observes the famous Landau-Pomeranchuk effect
\cite{LP}).

A different situation exists in the bremsstrahlung process at the
collision of electron and electron (positron) in colliding beams
experiments. The point is that the external factors act
differently on the radiating particle and on the recoil particle.
For the radiating particle the criterion of influence of external
factors is the same both at an electron scattering from a nucleus
and at a collision of particles. For the recoil particle the
effect turns out to be enhanced by the factor $\varepsilon^2/m^2$,
which is due to the fact that the main contribution to the
bremsstrahlung cross section give emitted by the recoil particle
virtual photons with very low energy
\begin{equation}
q_0 \sim \frac{m^2\omega}{\varepsilon (\varepsilon-\omega)},
\label{1.1}
\end{equation}
so that the formation length of virtual photon is
\begin{equation}
L_v(\omega)=l_f(q_0)= \frac{4\varepsilon^3
(\varepsilon-\omega)}{m^4\omega}. \label{1.2}
\end{equation}
This means that the effect for the recoil particles appears much
earlier than for the radiating particles. For example, the
Landau-Pomeranchuk effect distorted the whole bremsstrahlung
spectrum in a TeV range (for heavy elements) while it turns out
that the action on the recoil particle can be important for
contemporary colliding beam facilities in GeV range \cite{BK1}.

There are a few factors which could act on the recoil electron.
One of them is the presence of an external magnetic field in the
region of collision of particles \cite{BK1}-\cite{KS2}. If the
formation length of virtual photon $L_v$ turns out to be larger
than the formation length $l_H(\omega)$ of a photon with energy
$\omega$ in a magnetic field $H$  than the magnetic field will
limit the region of minimal momentum transfers, which will lead
to a decrease of the bremsstrahlung cross section and a change of
its spectrum. Another effect can appear due to the smallness of
the linear interval $l$ where the collision occurs in comparison
with $L_v(\omega)$ (see (\ref{1.2})). This was pointed out in
\cite{KS3}.

In the experiment \cite{exp1} devoted to study of the
bremsstrahlung spectrum $d\sigma_{\gamma}(\omega)$ carried out in
the electron-positron colliding beam facility VEPP-4 of Institute
of Nuclear Physics at an energy $\varepsilon=1.84~$GeV, a
deviation of the bremsstrahlung spectrum from the standard QED
spectrum was observed. This was attributed to the smallness of
the transverse size of the colliding beams. Theoretically the
problem of finite transverse dimensions was investigated in
\cite{BKS1} were the bremsstrahlung spectrum at $e^+e^-$ collision
have was calculated to within the power accuracy (the neglected
terms are of the order $1/\gamma=m/\varepsilon$) . Later the
problem was analyzed in \cite{BD}, \cite{KPS}, \cite{KSS} where
the found bremsstrahlung spectra coincide with obtained in
\cite{BKS1}.

It should be noted that in \cite{BKS1} (as well as in all other
papers mentioned above) an incomplete expression for the
bremsstrahlung spectrum was calculated. One has to perform the
subtraction associated with the extraction of pure fluctuation
process. Let us discuss this item in some details. The momentum
transfer ${\bf q}$ at collision is important for the radiation
process (the cross section contains factor ${\bf q}^2$ at ${\bf
q}^2 \ll m^2$). At the beam collision the momentum transfer may
arise due to interaction of the emitting particle with the
opposite beam as a whole (due to coherent interaction with
averaged field of the beam) and due to interaction with an
individual particle of the opposite beam. Here we are considering
{\it the incoherent} process only (connected with the incoherent
fluctuation of density) and so we have to subtract the coherent
contribution. The expression for the bremsstrahlung spectrum
found in \cite{BKS1} contains the mean value $<{\bf q}^2>$, while
the coherent contribution contains $<{\bf q}>^2$ and this term
has to be subtracted. We encountered with an analogous problem in
analysis of incoherent processes in the oriented crystals
\cite{BKS2} where it was pointed out (see p.407) that the
subtraction has to be done in the spectrum calculated in
\cite{BKS1}. Without the subtraction the results for the
incoherent processes in oriented crystals would be qualitatively
erroneous.

In Sec.2 a qualitative analysis of the incoherent radiation
process is given. In Sec.3 the general formulas for the
spectral-angular distributions of incoherent bremsstrahlung are
derived. The incoherent bremsstrahlung spectrum for the Gaussian
beams is calculated in Sec.4 in the form of double integrals. In
specific case of narrow beams (the size of beam is much smaller
than the characteristic impact parameter) the formulas are
simplified essentially (Sec.5). The experimental studies of
effect were performed with flat beams (the beam vertical size is
much smaller than horizontal one). This specific case is analyzed
in Sec.6, while comparison with data is given in Sec.7. In Sec.8
the possible application to the linear $e^+e^-$ collider tuning is
discussed.

\section{General analysis of probability of incoherent radiation}
\setcounter{equation}{0}

In this section we discuss in detail the conditions under which we
consider the incoherent radiation. One can calculate the photon
emission probability in the target rest frame, since the entering
combinations $\omega/\varepsilon$ and $\gamma\vartheta$ ($\gamma$
is the Lorentz factor $\gamma=\varepsilon/m$, $\vartheta$ is the
angle of photon emission) are invariant (within a relativistic
accuracy) and a transfer to any frame is elementary. We use the
operator quasiclassical method \cite{BK2}, \cite{BKS3}. Within
this method the photon formation length (time) is
\begin{eqnarray}
&& l_f=\frac{\varepsilon'}{\varepsilon
kv}=\frac{\varepsilon'}{\varepsilon \omega(1-\bf{n v})}\simeq
\frac{l_{f0}}{\zeta};
\nonumber \\
&& l_{f0}=\frac{1}{q_{min}}=\frac{2\varepsilon
\varepsilon'}{\omega m^2} = \frac{4\varepsilon' \gamma_c
\varepsilon_r}{\omega m^2};
\nonumber \\
&& \zeta = 1+\gamma^2 \vartheta^2, \quad \varepsilon'=\varepsilon
-\omega,
\label{2.1}
\end{eqnarray}
where $p_{\mu}= \varepsilon v_{\mu}~(v_{\mu}=(1, {\bf v}))$ is the
4-momentum of radiating particle, $\gamma_c=\varepsilon_c/m_c$,
$\varepsilon_c$ is the energy of target particle in the
laboratory frame, $m_c$ is its mass, $\varepsilon_r$ is the
energy of radiating particle in the laboratory frame;
~$k_{\mu}=(\omega,~\omega {\bf n}))$ is the photon 4-momentum,
$\vartheta$~ is the angle between vectors {\bf n} and {\bf v}.

In the case when the transverse dimension of beam $\sigma$ is
$\sigma \gg l_{f0}$ the impact parameters $\varrho \leq
\varrho_{max}=l_{f0}$ contribute. One can put that the particle
density in the target beam is a constant, so that the standard QED
formulas are valid. Note that the value $\varrho_{max}$ is the
relativistic invariant which is defined by the minimal value of
square of the invariant mass of the intermediate photon $|q^2|$.
In the case when the characteristic size of beams is smaller the
value $\varrho_{max}$ the lower value of $|q^2|$ is defined by
this size.

In the target rest frame the scattering length of emitting
particle is of order of the impact parameter $\varrho$. This
length is much smaller than the longitudinal dimension of the
target $\gamma_c l$ ($l$ is the length of target beam in the
laboratory frame). So one can neglect a variation of
configuration of the beam during the scattering time. A possible
variation of particle configuration in the beam during a long time
one can take into account in the adiabatic approximation.

Another limitation is connected with the influence of value of
transverse momentum arising from the electromagnetic field
$E=|{\bf E}|$ of colliding (target) beam on the photon formation
length. This value should be smaller than the characteristic
transverse momentum transfer $m\sqrt{\zeta}$ in the photon
emission process:
\begin{eqnarray}
&& \frac{e E l_f}{m\sqrt{\zeta}} \sim \frac{\alpha
N_c}{(\sigma_z+\sigma_y)l \gamma_c} \frac{1}{m\sqrt{\zeta}}
\frac{4\varepsilon' \gamma_c \varepsilon_r}{\omega \zeta m^2}
\nonumber \\
&& \sim \frac{2 \alpha N_c}{(\sigma_z+\sigma_y)l}
\frac{1}{m\sqrt{\zeta}} \frac{2\varepsilon' \varepsilon_r}{\omega
\zeta m^2}=\frac{4 \alpha N_c \gamma_r \lambda_c^2
\varepsilon'}{(\sigma_z+\sigma_y)l \zeta^{3/2} \omega} \ll 1,
\label{2.2}
\end{eqnarray}
here $\alpha = 1/137$, $N_c$ is the number of particles in the
target beam, $\sigma_z$ and $\sigma_y$ are the vertical and
horizontal transverse dimensions of target beam. Note that the
ratio $\gamma/l$ is the relativistic invariant. This condition
can be presented in invariant form
\begin{equation}
\frac{2 \chi}{u \zeta^{3/2}} \ll 1,
\label{2.3}
\end{equation}
where $\displaystyle{\chi=\frac{\gamma }{E_0}|{\bf
E}_{\perp}+{\bf v}\times {\bf H}|,~
u=\frac{\omega}{\varepsilon'},~ E_0=\frac{m^2}{e}=1.32 \cdot
10^{16} {\rm V/cm}}$. Since the main contribution to the spectral
probability of radiation gives angles $\vartheta \sim 1/\gamma$
($\zeta \sim 1$) this condition takes the form $\chi/u \ll 1$.
For the case $\chi/u \gg 1$ the condition (\ref{2.3}) can be
satisfied for the large photon emission angles $\zeta \simeq
\gamma^2\vartheta^2> (\chi/u)^{2/3} \gg 1$. Under these conditions
the formation length $l_{f}=l_{f0}/\zeta$ decreases as
$(\chi/u)^{2/3}$. The same inhibition factor acquires the
bremsstrahlung probability \cite{BKS4}.

We consider now the spectral distribution of radiation
probability in the case $\chi \ll 1$ (this condition is fulfilled
in all existing installations), so
\begin{equation}
\chi \sim \alpha N_c \gamma
\frac{\lambda_c^2}{(\sigma_z+\sigma_y)l} \ll 1. \label{2.4}
\end{equation}
Only the soft photons $(\omega \leq \chi \varepsilon \ll
\varepsilon)$ contribute to the {\it coherent} radiation
("beamstrahlung") while the hard photon region $\omega \gg \chi
\varepsilon$ is suppressed exponentially as it is known from the
classical radiation theory. As it was mentioned in the soft photon
region $(\omega \leq \chi \varepsilon \ll \varepsilon)$, the
spectral probability of {\it bremsstrahlung} is suppressed by the
factor $(\omega/\varepsilon\chi)^{2/3}$ only.  On the contrary,
the spectral probability of the bremsstrahlung is negligible
comparing with the beamstrahlung taking into consideration that
the mean square of multiple scattering angle during all time of
beam collisions is small comparing with the value $1/\gamma^2$:
\begin{equation}
\gamma^2 <\vartheta_s^2>=\frac{<q_s^2>}{m^2} \simeq
\frac{8\alpha^2N_c \lambda_c^2}{\sigma_z\sigma_y}L \ll 1,
\label{2.5}
\end{equation}
where $L$ is the characteristic logarithm of scattering problem
(in the typical experimental condition $L \sim 10$).

It was supposed in the above estimations of beamstrahlung
probability that the radiation formation length is shorter than
the target beam length
\begin{equation}
\frac{l_f}{l} \sim
\frac{1}{u}\left(1+\frac{\chi}{u}\right)^{-2/3}\frac{\gamma
\lambda_c }{l} < 1.
\label{2.6}
\end{equation}
Besides it was supposed that one can neglect a variation of the
impact parameter $\mbox{\boldmath$\varrho$}$ and therefore of the
transverse electric field ${\bf
E}_{\perp}(\mbox{\boldmath$\varrho$})$ during the beam collision.
This is true when disruption parameter is enough small
\begin{equation}
D_i = \frac{2\alpha N_c \lambda_c l}{\gamma_r \sigma_i
(\sigma_z+\sigma_y)} \ll 1,\quad (i=z,y)
\label{2.7}
\end{equation}

So, we consider the incoherent bremsstrahlung under following
conditions:
\begin{equation}
\chi \ll 1,\quad \frac{\chi}{u} \ll 1,\quad D_i \ll 1. \label{2.8}
\end{equation}

\section{{Spectral-angular distribution of the}
\newline {incoherent bremsstrahlung probability}}

\setcounter{equation}{0}

In this section we derive the basic expression for the incoherent
bremsstrahlung probability at collision of two beams with bounded
transverse dimensions.

We consider first the photon emission at collision of an electron
with one particle with the transverse coordinate {\bf x}. We
select the impact parameter
$\varrho_0=|\mbox{\boldmath$\varrho$}_0|$ which is small comparing
with the typical transverse beam dimension $\sigma$ but which is
large comparing with the electron Compton length $\lambda_c$
($\lambda_c \ll \varrho_0 \ll \sigma$). In the interval of impact
parameters  $\varrho = |{\bf r}_\perp -{\bf x}| \geq \varrho_0$,
where ${\bf r}_\perp$ is the transverse coordinate of emitting
electron, the probability of radiation summed over the momenta of
final particle can be calculated using the classical trajectory
of particle. Indeed,one can neglect by the value of commutators
$|[\hat{p}_{\perp i}, \varrho_j]|=\delta_{ij}$ comparing with the
value $p_{\perp} \varrho$ in this interval ($p_{\perp} \varrho
\geq m\varrho_0 \gg 1$). In this case the expression for the
probability has the form (see \cite{BKS3}, Eqs.(7.3) and (7.4))
\begin{equation}
dw = |M(\mbox{\boldmath$\varrho$})|^2w_r({\bf r_{\perp}})d^2
r_{\perp} d^3k,
\label{3.1}
\end{equation}
where
\begin{equation}
M(\mbox{\boldmath$\varrho$})=\frac{e}{2\pi\sqrt{\omega}}\int_{-\infty}^{\infty}
R(t)\exp(ik'x(t))dt,\quad k'=\frac{\varepsilon}{\varepsilon'}k.
\label{3.2}
\end{equation}
Here $w_r({\bf r_{\perp}})d^2 r_{\perp}$ is the probability to
find the emitting particle with the impact parameter
$\mbox{\boldmath$\varrho$} = {\bf r}_\perp -{\bf x}$ in the
interval $d^2 \varrho = d^2 r_{\perp},~ R(t)=R({\bf p}(t)),~
kx(t)=\omega t-{\bf kr}(t)$ (for details see \cite{BKS3}, Sec.
7.1). Integrating by parts in the last equation and taking into
account that $|{\bf q}_{\perp}(\mbox{\boldmath$\varrho$})| \leq
1/\varrho_0 \ll m$ we find
\begin{equation}
M(\mbox{\boldmath$\varrho$}) \simeq \frac{i e}{2\pi
\sqrt{\omega}}\int_{-\infty}^{\infty}
\exp(ik'vt)\frac{d}{dt}~\frac{R(t)}{k'v(t)}dt \simeq \frac{i
e}{2\pi \sqrt{\omega}} {\bf m}(\mbox{\boldmath$\varrho$})
\frac{\partial}{\partial {\bf p_{\perp}}} \frac{R({\bf
p_{\perp}})}{k'v}, \label{3.3}
\end{equation}
where
\begin{eqnarray}
&&{\bf p}_{\perp}={\bf p}-{\bf n}({\bf n}{\bf p}) \simeq
\varepsilon ({\bf v}-{\bf n}),
\nonumber \\
&& {\bf m}(\mbox{\boldmath$\varrho$}) = \int_{-\infty}^{\infty}
\exp(ik'vt)\dot{{\bf q}}(\mbox{\boldmath$\varrho$}, t) dt =
-\frac{\partial}{\partial \mbox{\boldmath$\varrho$}}
\int_{-\infty}^{\infty}
\exp\left(\frac{it}{l_f}\right)V(\sqrt{\varrho^2+t^2})dt
\nonumber \\
&& = \frac{2\alpha}{l_f}
\frac{\mbox{\boldmath$\varrho$}}{\varrho}K_1(\frac{\varrho}{l_f})=2\alpha
q_{min} \zeta K_1(\varrho q_{min} \zeta)
\frac{\mbox{\boldmath$\varrho$}}{\varrho},
\label{3.4}
\end{eqnarray}
for the Coulomb potential, $K_1(z)$ is the modified Bessel
function (the Macdonald function), $R({\bf p_{\perp}})$ has a form
of matrix element for the free particles:
\begin{eqnarray}
&& R({\bf
p_{\perp}})=\varphi_{s'}^{+}(A+i\mbox{\boldmath$\sigma$}{\bf
B})\varphi_{s}; \quad A \simeq
\frac{m(\varepsilon+\varepsilon')}{2\varepsilon
\varepsilon'}({\bf e}^{\ast}{\bf u}),
\nonumber \\
&& {\bf B} \simeq \frac{m\omega}{2\varepsilon \varepsilon'}({\bf
e}^{\ast}\times ({\bf n}-{\bf u})); \quad {\bf u}=\frac{{\bf
p}_{\perp}}{m},~\zeta=1+{\bf u}^2,~k'v=q_{min}\zeta,
\label{3.5}
\end{eqnarray}
where the vector {\bf e} describes the photon polarization and
the spinors $\varphi_s$ and $\varphi_{s'}$ describe the
polarization of the initial and final electrons correspondingly.

In the interval of impact parameters $\varrho \leq \lambda_c$ the
expectation value of operator
$<\mbox{\boldmath$\varrho$}|M^{+}M|\mbox{\boldmath$\varrho$}>$
cannot be written in the form (\ref{3.1}) since the entering
operators become noncommutative inside the expectation value.
However because of the condition $\lambda_c \ll \sigma$ in this
interval $w_r({\bf r}_{\perp}) \simeq w_r({\bf x}) +
O(\frac{\lambda_c}{\sigma})$ one can neglect effect of the
inhomogeneous distribution. For the same reason in the
calculation of correction to the probability of photon emission,
which is defined as the difference of $dw(\sigma)$ and the
probability of photon emission in a inhomogeneous medium, one can
extend the integration interval into region $\varrho \leq
\varrho_0$.

In this paper we consider the incoherent bremsstrahlung which can
be considered as the photon emission due to fluctuations of the
potential $V$ connected with uncertainty of a particle position in
the transverse to its momentum plane. Because of this we have to
calculate the dispersion of the vector ${\bf
m}(\mbox{\boldmath$\varrho$})$ with respect to the transverse
coordinate $\mbox{\boldmath$\varrho$}$:
\begin{eqnarray}
&& <m_i m_j> - <m_i>< m_j>= \int m_i({\bf r}_\perp -{\bf x})
m_j({\bf r}_\perp -{\bf x}) w_c({\bf x})d^2x
\nonumber \\
&& - \int m_i({\bf r}_\perp -{\bf x})w_c({\bf x})d^2x \int
m_j({\bf r}_\perp -{\bf x})w_c({\bf x})d^2x,
\label{3.6}
\end{eqnarray}
where $w_c({\bf x})$ is the distribution function of target
particles normalized to the unity.

As a result we obtain the following expression for the correction
to the probability of photon emission connected with the
restricted transverse dimensions of colliding beams of charged
particles:
\begin{eqnarray}
&& dw_1=\frac{\alpha}{(2\pi)^2}\frac{d^3k}{\omega}T_{ij}({\bf e},
{\bf p}_{\perp}, s, s')L_{ij}, \quad
T_{ij}=\left[\frac{\partial}{\partial p_{\perp
i}}\frac{R^{\ast}({\bf
p}_{\perp})}{k'v}\right]\left[\frac{\partial}{\partial p_{\perp
j}}\frac{R^{\ast}({\bf p}_{\perp})}{k'v}\right],
\nonumber \\
&& L_{ij}=\int m_i(\mbox{\boldmath$\varrho$})
m_j(\mbox{\boldmath$\varrho$})\left(w_r({\bf
x}+\mbox{\boldmath$\varrho$})-w_r({\bf x})\right)w_c({\bf
x})d^2xd^2\varrho
\nonumber \\
&&- \left(\int m_i(\mbox{\boldmath$\varrho$})w_c({\bf
x}-\mbox{\boldmath$\varrho$})d^2\varrho \right)\left(\int
m_j(\mbox{\boldmath$\varrho$})w_c({\bf
x}-\mbox{\boldmath$\varrho$})d^2\varrho \right)w_r({\bf x })d^2x.
\label{3.7}
\end{eqnarray}

Averaging over the polarization of initial electrons and summing
over the polarization of final electrons we find
\begin{equation}
 T_{ij}=\frac{l_f}{\varepsilon\varepsilon'}\Big[e_i e_j-
\frac{2{\bf e u}}{\zeta}(e_i u_j + u_i e_j) + \frac{4({\bf e
u})^2}{\zeta^2} u_i u_j+
\frac{\omega^2}{4\varepsilon\varepsilon'}\delta_{ij}\Big].
\label{3.8}
\end{equation}
Note, that one can choose the real vector {\bf e} since only the
linear polarization could arise in the case of unpolarized
electrons.

After summation in (\ref{3.8}) over the polarization of emitted
photon we have
\begin{equation}
 T_{ij}=\frac{l_f}{2\varepsilon\varepsilon'}\left(v \delta_{ij}-
 \frac{8}{\zeta^2}u_i u_j\right),\quad v=\frac{\varepsilon}{\varepsilon'}
 +\frac{\varepsilon'}{\varepsilon},\quad \zeta = 1+\gamma^2 \vartheta^2.
\label{3.9}
\end{equation}

Finally, averaging the last expression over the azimuth angle of
emitted photon we obtain
\begin{equation}
T_{ij}=\frac{l_f}{2\varepsilon\varepsilon'}U(\zeta)
 \delta_{ij},\quad
U(\zeta)=v-\frac{4(\zeta-1)}{\zeta^2}.
\label{3.10}
\end{equation}

Substituting the expression obtained into (\ref{3.7}) we find the
correction to the probability of photon emission connected with
the restricted transverse dimensions of colliding beams of charged
particles:
\begin{equation}
dw_1=\frac{\alpha^3}{\pi
m^2}\frac{\varepsilon'}{\varepsilon}\frac{d
\omega}{\omega}U(\zeta) F(\omega, \zeta) d\zeta,
\label{3.11}
\end{equation}
where
\begin{eqnarray}
&& F(\omega, \zeta)=F^{(1)}(\omega, \zeta)-F^{(2)}(\omega,
\zeta),
\nonumber \\
&& F^{(1)}(\omega, \zeta)=\frac{2\eta^2}{\zeta^2}\int K_1^2(\eta
\varrho)\left(w_r({\bf x}+\mbox{\boldmath$\varrho$})-w_r({\bf
x})\right)w_c({\bf x})d^2xd^2\varrho,
\nonumber \\
&& F^{(2)}(\omega, \zeta)=\frac{2\eta^2}{\zeta^2}\int\left(\int
K_1(\eta \varrho)\frac{\mbox{\boldmath$\varrho$}}{\varrho}w_c({\bf
x}-\mbox{\boldmath$\varrho$})d^2\varrho \right)^2w_r({\bf x})d^2x,
\label{3.12}
\end{eqnarray}
here $\eta=q_{min}\zeta$.

Using the integral
\begin{equation}
\int K_1^2(\eta \varrho) \varrho d\varrho =
\frac{\varrho^2}{2}\left[K_1^2(\eta \varrho)-K_0(\eta
\varrho)K_2(\eta \varrho)\right]
\label{3.13}
\end{equation}
and integrating by parts we obtain
\begin{eqnarray}
&& F(\omega, \zeta)= \frac{\eta^2}{\zeta^2}\Big[\int
\left[K_0(\eta \varrho)K_2(\eta \varrho)-K_1^2(\eta
\varrho)\right]\varrho
\frac{d\Phi(\mbox{\boldmath$\varrho$})}{d\varrho}d^2\varrho
\nonumber \\
&& -2\int\left(\int K_1(\eta
\varrho)\frac{\mbox{\boldmath$\varrho$}}{\varrho}w_c({\bf
x}-\mbox{\boldmath$\varrho$})d^2\varrho \right)^2w_r({\bf
x})d^2x\Big],
\label{3.14}
\end{eqnarray}
where
\begin{equation}
\Phi(\mbox{\boldmath$\varrho$})=\int w_r({\bf
x}+\mbox{\boldmath$\varrho$}) w_c({\bf x})d^2x.
\label{3.15}
\end{equation}

In the general case the axes of colliding beams are displaced
with respect each other in the transverse plane by the vector
${\bf x}_0$ with coordinates $z_0,y_0$. In this case we have to
consider
\begin{equation}
w_r({\bf x}) \rightarrow w_r({\bf x}+{\bf
x}_0),~F^{(1,2)}(\omega, \zeta) \rightarrow F^{(1,2)}(\omega,
\zeta, {\bf x}_0),~ \Phi(\mbox{\boldmath$\varrho$}) \rightarrow
\Phi(\mbox{\boldmath$\varrho$}+{\bf x}_0)
\label{3.15a}
\end{equation}

The first term in the expression for $F(\omega, \zeta)$ in
(\ref{3.14}) coincides with the function $F(\omega, \zeta)$
defined in \cite{BKS1}, Eq.(13). The second (subtraction) term in
(\ref{3.14})which is naturally arisen in this derivation was
missed in Eq.(13), \cite{BKS1}  as it was said above. The
expression (\ref{3.11}) is consistent with Eq.(21.6) in the book
\cite{BKS3} (see also Eq.(2.2) in \cite{BKS2}) where another
physical problem was analyzed. It is the incoherent
bremsstrahlung in the oriented crystals.

Below we restrict ourselves to the case of unpolarized electrons
and photons. Influence of bounded transverse size on the
probability of process with polarized particles will be considered
elsewhere.

\section{Gaussian beams}

\setcounter{equation}{0}

For calculation of explicit expression for the bremsstrahlung
cross section we have to specify the distributions of particles
in the colliding beams. Here we consider the actual case of
Gaussian beams. Using the Fourier transform we have
\begin{eqnarray}
\hspace{-10mm}&& w({\bf x})=\frac{1}{(2\pi)^2}\int d^2q
\exp(-i{\bf q x})w(\bf q);
\nonumber \\
\hspace{-10mm}&& w_r({\bf
q})=\exp\left[-\frac{1}{2}(q_z^2\Delta_z^2+q_y^2\Delta_y^2)\right],~
w_c({\bf
q})=\exp\left[-\frac{1}{2}(q_z^2\sigma_z^2+q_y^2\sigma_y^2)\right],
\label{4.1}
\end{eqnarray}
where as above the index $r$ relates to the radiating beam and the
index $c$ relates to the target beam, $\Delta_z$ and $\Delta_y$
($\sigma_z$ and $\sigma_y$) are the vertical and horizontal
transverse dimensions of radiating (target) beam. Substituting
(\ref{4.1}) into Eq.(\ref{3.15}) we find
\begin{eqnarray}
\hspace{-5mm}&& \Phi(\mbox{\boldmath$\varrho$})=
\frac{1}{(2\pi)^2}\int d^2q \exp(-i{\bf q}
\mbox{\boldmath$\varrho$})\exp\left[-\frac{q_z^2}{4\Sigma_z^2}
-\frac{q_y^2}{4\Sigma_y^2}\right]=
\frac{\Sigma_z\Sigma_y}{\pi}\exp[-\varrho_z^2\Sigma_z^2-\varrho_y^2\Sigma_y^2];
\nonumber \\
\hspace{-5mm}&& \Sigma_z^2=\frac{1}{2(\sigma_z^2+\Delta_z^2)},~
\Sigma_y^2=\frac{1}{2(\sigma_y^2+\Delta_y^2)},
\label{4.2}
\end{eqnarray}

Below we consider the general situation when the axes of
colliding beams are displaced with respect each other in the
transverse plane by the vector ${\bf x}_0$ with the coordinates
$z_0,y_0$. This displacement has essential influence on the
luminosity. For the processes for which the short distances are
essential only (e.g. double bremsstrahlung \cite{BK1}) the
probability of process is the product of the cross section and
luminosity. The geometrical luminosity per bunch, not taking into
account the disruption effects, is given by
\begin{equation}
{\cal L}=N_cN_r \Phi({\bf x}_0),
\label{4.3}
\end{equation}
where as above $N_r$ and $N_c$ are the number of particles in the
radiating and target beams correspondingly. We will use the same
definition for our case. Then we have
\begin{equation}
dw_{\gamma}= \Phi({\bf x}_0)d\sigma_{\gamma} ,\quad d\sigma_1=
\Phi^{-1}({\bf x}_0) dw_1, \label{4.4}
\end{equation}
where $dw_1$ is defined in Eq.(\ref{3.11}).

We calculate first the function $F^{(1)}(\omega, \zeta)$ in
Eq.(\ref{3.12}) for the case of coaxial  beams when ${\bf x}_0=0$.
Passing on to the momentum representation with the help of
formula (\ref{4.1}) we find
\begin{equation}
F^{(1)}(\omega, \zeta)= - \frac{1}{2\pi\zeta^2}\int w_r({\bf q})
w_c({\bf q}) F_2\left(\frac{q}{2\eta}\right) qdq d\varphi,
\label{4.5}
\end{equation}
where $\eta=q_{min}\zeta$ is introduced in (\ref{3.12}),
\begin{eqnarray}
&& F_2\left(\frac{q}{2\eta}\right) = \frac{\eta^2}{\pi}\int
K_1^2(\eta \varrho)(1-\exp(-i{\bf q}
\mbox{\boldmath$\varrho$}))d^2\varrho,
\nonumber \\
&& F_2(x)=\frac{2x^2+1}{x\sqrt{1+x^2}}\ln(x+\sqrt{1+x^2})-1,\quad
q_{min}=m^3\omega/4\varepsilon^2 \varepsilon',
\label{4.6}
\end{eqnarray}
here value $q_{min}$ is defined in c.m.s. of colliding particles.
The function $F_2(x)$ encounters in the radiation theory. To
calculate the corresponding contribution into the radiation
spectrum we have to substitute (\ref{4.5}) into Eq.(\ref{3.11})
and take the integrals. After substitution of variable in
(\ref{4.5})
\begin{equation}
w=\frac{q}{2q_{min}\zeta},
\label{4.7}
\end{equation}
we obtain the integral over $\zeta$ in Eq.(\ref{3.11}):
\begin{eqnarray}
&&\int_{1}^{\infty} \left( v -\frac{4}{\zeta} +
\frac{4}{\zeta^2}\right)\exp(-s^2\zeta^2)d\zeta \equiv f(s)
\nonumber \\
&& = \frac{\sqrt{\pi}}{2s}(v-8s^2){\rm
Erfc}(s)+4\displaystyle{e^{-s^2}}+2{\rm Ei}(-s^2),
\label{4.8}
\end{eqnarray}
where
\begin{equation}
s=w r q_{min}, \quad r^2= \Sigma_z^{-2}\cos^2\varphi+
\Sigma_y^{-2}\sin^2\varphi
\label{4.8a}
\end{equation}
Making use of Eq.(\ref{4.4}) we find for the spectrum
\begin{eqnarray}
&& d\sigma_1^{(1)}=\frac{2\alpha^3}{m^2}
\frac{\varepsilon'}{\varepsilon}
\frac{d\omega}{\omega}f^{(1)}(\omega),
\nonumber \\
&& f^{(1)}(\omega)= -\frac{1}{\pi\Sigma_z
\Sigma_y}\int_{0}^{2\pi}\frac{d\varphi}{\Sigma_z^{-2}\cos^2\varphi+
\Sigma_y^{-2}\sin^2\varphi} \int_{0}^{\infty}F_2(z)f(s)s ds,
\nonumber \\
&& z^2=\frac{s^2}{q_{min}^2}\frac{1}{\Sigma_z^{-2}\cos^2\varphi+
\Sigma_y^{-2}\sin^2\varphi}.
\label{4.9}
\end{eqnarray}
This formula is quite convenient for the numerical calculations.

In the case ${\bf x}_0 \neq 0$ we will use straightforwardly
Eqs.(\ref{3.12}) and (\ref{4.4}). Taking into account (\ref{4.2})
we have for the difference
\begin{eqnarray}
\hspace{-15mm}&& \Delta^{(1)}({\bf x}_0) \equiv
\frac{1}{2\pi}(\Phi^{-1}({\bf x}_0)F^{(1)}({\bf x}_0)-\Phi^{-1}(0)
F^{(1)}(0))
\nonumber \\
\hspace{-15mm}&&=\frac{\eta^2}{\pi\zeta^2}\int
K_1^2(\eta\varrho)\exp\left[-\varrho_z^2\Sigma_z^2-
\varrho_y^2\Sigma_y^2\right] \{\exp\left[-2\varrho_zz_0\Sigma_z^2-
2\varrho_yy_0\Sigma_y^2\right]-1\}d^2\varrho,
 \label{4.10}
\end{eqnarray}
where the function $F^{(1)}({\bf x}_0)$ is defined in
Eqs.(\ref{3.12}), (\ref{3.15a}).

Using the Macdonald's formula (see e.g.\cite{BE}, p.53)
\begin{equation}
2K_1^2(\eta\varrho)=\int_{0}^{\infty} \exp\left[ -\varrho^2
t-\frac{\eta^2}{2t} \right]K_1\left(
\frac{\eta^2}{2t}\right)\frac{dt}{t} \label{4.11}
\end{equation}
and taking the Gaussian integrals over $\varrho_z$ and $\varrho_y$
we get
\begin{equation}
 \Delta^{(1)}({\bf x}_0) =\frac{1}{\zeta^2}\int_{0}^{\infty}
\frac{\exp\left( -\frac{\eta^2}{2t}\right)K_1\left(
\frac{\eta^2}{2t}\right)}{\sqrt{t+\Sigma_z^2}\sqrt{t+\Sigma_y^2}}
\left\{ \exp\left[ \frac{z_0^2\Sigma_z^4}{t+\Sigma_z^2}+
\frac{y_0^2\Sigma_y^4}{t+\Sigma_y^2}\right]-1
\right\}\frac{\eta^2 dt}{2t}. \label{4.12}
\end{equation}
For the correction to the cross section (see Eqs.(\ref{4.4}) and
(\ref{4.9})) we have correspondingly
\begin{equation}
 d\sigma_1^{(1)}=\frac{2\alpha^3}{m^2}
\frac{\varepsilon'}{\varepsilon}
\frac{d\omega}{\omega}[f^{(1)}(\omega)+J^{(1)}(\omega, {\bf x}_0
)],
\label{4.13}
\end{equation}
where
\begin{equation}
 J^{(1)}(\omega, {\bf x}_0)= \int_{1}^{\infty} U(\zeta)
 \Delta^{(1)}({\bf x}_0)d\zeta.
\label{4.14}
\end{equation}

Now we pass over to the calculation of the second (subtraction)
term $F^{(2)}(\omega, \zeta)$ in Eq.(\ref{3.12}). Using
Eq.(\ref{4.1}) we get
\begin{equation}
{\bf I}= \eta \int K_1(\eta
\varrho)\frac{\mbox{\boldmath$\varrho$}}{\varrho}w_c({\bf
x}-\mbox{\boldmath$\varrho$})d^2\varrho =
\frac{\eta}{(2\pi)^2}\int S(q) \frac{{\bf q}}{q}\exp(-i{\bf q
x})w_c({\bf q})d^2q, \label{4.15}
\end{equation}
where
\begin{eqnarray}
&& S(q)=  \int K_1(\eta\varrho)\frac{{\bf q
}\mbox{\boldmath$\varrho$}}{q\varrho}\exp(i{\bf q
}\mbox{\boldmath$\varrho$})d^2\varrho
\nonumber \\
&& = 2\pi i \int K_1(\eta\varrho) J_1(q\varrho) \varrho d\varrho
=  2\pi i~ \frac{q}{\eta} \frac{1}{q^2+\eta^2}.
\label{4.16}
\end{eqnarray}
Using the exponential parametrization
\begin{equation}
\frac{1}{q^2+\eta^2}=\frac{1}{4}\int_{0}^{\infty}
\exp\left[-\frac{s}{4}(q^2+\eta^2) \right]ds
\label{4.17}
\end{equation}
and taking the Gaussian integrals over $q_z$ and $q_y$ we obtain
\begin{eqnarray}
&& {\bf I}= \int_{0}^{\infty} \exp\left[-\frac{\eta^2 s}{4} -
\frac{z^2}{s+2\sigma_z^2} - \frac{y^2}{s+2\sigma_y^2} \right]
\nonumber \\
&& \times \left[ \frac{z {\bf e}_z}{s+2\sigma_z^2}+ \frac{y {\bf
e}_y}{s+2\sigma_y^2}\right]
\frac{ds}{\sqrt{s+2\sigma_z^2}\sqrt{s+2\sigma_y^2}}, \label{4.18}
\end{eqnarray}
where ${\bf e}_z$ and ${\bf e}_y$ are the unit vectors along axes
$z$ and $y$. Substituting (\ref{4.18}) into Eq.(\ref{3.12}),
taking the Gaussian integrals over $z$ and $y$ and using
Eq.(\ref{4.4}) we get the correction to the cross section
\begin{equation}
d\sigma_1^{(2)}=-\frac{2\alpha^3}{m^2}
\frac{\varepsilon'}{\varepsilon}
\frac{d\omega}{\omega}J^{(2)}(\omega, {\bf x}_0), \label{4.19}
\end{equation}
where
\begin{eqnarray}
&& J^{(2)}(\omega, {\bf x}_0)= \frac{\sqrt{a b}}{\Sigma_z
\Sigma_y} \exp(z_0^2 \Sigma_z^2+y_0^2 \Sigma_y^2)\int_{0}^{\infty}
ds_1 \int_{0}^{\infty} ds_2  g\left( \frac{q_{min}\sqrt{s}}{2}
\right)G(s_1, s_2, {\bf x}_0 ),
\nonumber \\
&& G(s_1, s_2, {\bf x}_0 )= \left( \frac{a_1a_2 b_1 b_2}{A
B}\right)^{1/2}\left[\frac{a_1a_2}{A}\left( \frac{1}{2}+
\frac{z_0^2a^2}{A}\right)+\frac{b_1b_2}{B}\left( \frac{1}{2}+
\frac{y_0^2b^2}{B}\right)\right]
\nonumber \\
&& \times \exp\left[-\frac{z_0^2a}{A}(a_1+a_2)
-\frac{y_0^2b}{B}(b_1+b_2) \right].
\label{4.20}
\end{eqnarray}
Here the function $g$ appears as a result of integration over
$\zeta$:
\begin{eqnarray}
\hspace{-15mm}&&g(q)=\int_{1}^{\infty}\left( v-\frac{4}{\zeta}+
\frac{4}{\zeta^2}\right)\exp(-q^2\zeta^2)\frac{d\zeta}{\zeta^2}=\left(
v -\frac{2}{3}\right)\exp(-q^2)
\nonumber \\
\hspace{-15mm}&&-2q^2 \int_{1}^{\infty}\left( v-\frac{2}{\zeta}+
\frac{4}{3\zeta^2}\right)\exp(-q^2\zeta^2)d\zeta
\nonumber \\
\hspace{-15mm}&&= \left( v -\frac{2}{3}\right)\exp(-q^2)
-2q^2\left[ \frac{\sqrt{\pi}}{2q}\left(v-\frac{8}{3}q^2\right){\rm
Erfc}(q)+\frac{4}{3}\displaystyle{e^{-q^2}}+{\rm Ei}(-q^2)\right].
\label{4.21}
\end{eqnarray}
In (\ref{4.20}) we introduced the following notations
\begin{eqnarray}
&& a=\frac{1}{2\Delta_z^2}, \quad b=\frac{1}{2\Delta_y^2}, \quad
a_{1,2}=\frac{1}{s_{1,2}+2\sigma_z^2},\quad
b_{1,2}=\frac{1}{s_{1,2}+2\sigma_y^2},
\nonumber \\
&& A=a_1+a_2+a, \quad B=b_1+b_2+b,\quad s=s_1+s_2.
\label{4.22}
\end{eqnarray}

\section{Narrow beams}

\setcounter{equation}{0}

This is the case when the ratio $q_{min}/(\Sigma_z+\Sigma_y) \ll
1$, so that the main contribution to the integral (\ref{4.9})
gives the region $q \sim \zeta \sim 1,~z \gg 1$. Using the
asymptotics of function $F_2(z)$ at $z \gg 1$
\begin{equation}
F_2(z) \simeq \ln (2z)^2-1
\label{5.1}
\end{equation}
and the following integrals
\begin{eqnarray}
&&\hspace{-10mm}
\frac{1}{2\pi\Sigma_z\Sigma_y}\int_{0}^{2\pi}\frac{d\varphi}
{\Sigma_z^{-2}\cos^2\varphi+\Sigma_y^{-2}\sin^2\varphi}=1,
\nonumber \\
&&\hspace{-10mm}
\frac{1}{2\pi\Sigma_z\Sigma_y}\int_{0}^{2\pi}\frac{d\varphi}
{\Sigma_z^{-2}\cos^2\varphi+\Sigma_y^{-2}\sin^2\varphi}\ln
\frac{4}{\Sigma_z^{-2}\cos^2\varphi+\Sigma_y^{-2}\sin^2\varphi}=\ln
(\Sigma_z+\Sigma_y)^2,
\nonumber \\
&&\hspace{-10mm}\int_{1}^{\infty}dq^2 \left(\alpha-\beta \ln
q^2\right)\int_{1}^{\infty}\left( v-\frac{4}{\zeta}+
\frac{4}{\zeta^2}\right)\exp(-q^2\zeta^2)d\zeta
\nonumber \\
&&
\hspace{-10mm}=\left(v-\frac{2}{3}\right)[\alpha+\beta(2+C)]+\frac{2}{9}\beta,
\label{5.2}
\end{eqnarray}
where C is Euler's constant $C = 0.577...$, we get for the
function $f^{(1)}(\omega)$ (\ref{4.13}) the following expression
\begin{equation}
f^{(1)}(\omega) \simeq \left(v-\frac{2}{3}\right)\left(2\ln
\frac{q_{min}}{\Sigma_z+\Sigma_y}+3+C \right) + \frac{2}{9},~\quad
q_{min} \ll (\Sigma_z+\Sigma_y). \label{5.3}
\end{equation}
This expression agrees with Eq.(24) of \cite{BKS1}.

Under the assumption used in (\ref{5.3}) and the additional
condition $q_{min}(z_0+y_0)\ll 1$ the main contribution to the
integral in (\ref{4.12}) gives the region $t \gg \eta^2$. In this
case one can use the asymptotic expansion $K_1(z) \simeq 1/z (z
\ll 1)$. Then we have for the function $J^{(1)}(\omega, {\bf
x}_0)$ in Eq.(\ref{4.13}) the following expression
\begin{eqnarray}
&& J^{(1)}(\omega, {\bf x}_0) \simeq \left(v-\frac{2}{3}\right)J
\nonumber \\
&&
J=\int_{0}^{\infty}\left[\exp\left(\frac{z_0^2\Sigma_z^4}{t+\Sigma_z^2}
+\frac{y_0^2\Sigma_y^4}{t+\Sigma_y^2}\right)-1\right]
\frac{dt}{\sqrt{t+\Sigma_z^2}\sqrt{t+\Sigma_y^2}}. \label{5.4}
\end{eqnarray}
The expression (\ref{5.4}) is consistent with Eq.(26) of
\cite{BKS1}.

In the case $({\bf x}_0^2+\sigma_z^2+\sigma_y^2)q_{min}^2 \ll 1$
the main contribution to the integral in (\ref{4.20}) gives the
interval $s q_{min}^2 \sim ({\bf
x}_0^2+\sigma_z^2+\sigma_y^2)q_{min}^2 \ll 1$. Keeping the main
term of expansion over $q^2$ in Eq.(\ref{4.21}) we get
\begin{equation}
g\left(\frac{q_{min}\sqrt{s}}{2}\right) \simeq v-\frac{2}{3}.
\label{5.5}
\end{equation}
The same result can be obtained if one neglects the term
containing $\eta^2$ in the exponent of integrand in
Eq.(\ref{4.18}).

Summing the cross section $d\sigma =
d\sigma_1^{(1)}+d\sigma_1^{(2)}$ with the standard QED
bremsstrahlung cross section
\begin{equation}
d\sigma_0= \frac{2\alpha^3}{m^2} \frac{\varepsilon'}{\varepsilon}
\frac{d\omega}{\omega}\left(v-\frac{2}{3}\right)\left(\ln
\frac{m^2}{q_{min}^2}-1 \right),
\label{5.6}
\end{equation}
we get the cross section for the case of interaction of narrow
beams
\begin{eqnarray}
&& d\sigma_{\gamma}= d\sigma_0 + d\sigma_1 =
\frac{2\alpha^3}{m^2} \frac{\varepsilon'}{\varepsilon}
\frac{d\omega}{\omega}\Bigg\{\left(v-\frac{2}{3}\right)\Bigg[2\ln
\frac{m}{\Sigma_z+\Sigma_y} +C +2
\nonumber \\
&& +J -J_{-}\Bigg]+\frac{2}{9}\Bigg\},\quad
v=\frac{\varepsilon}{\varepsilon'}+\frac{\varepsilon'}{\varepsilon},\quad
\varepsilon'=\varepsilon - \omega,
 \label{5.7}
\end{eqnarray}
where $J$ is given in (\ref{5.4}),
\begin{equation}
 J_{-}=\frac{\sqrt{a b}}{\Sigma_z
\Sigma_y} \exp(z_0^2 \Sigma_z^2+y_0^2 \Sigma_y^2)\int_{0}^{\infty}
ds_1 \int_{0}^{\infty} ds_2 G(s_1, s_2, {\bf x}_0 ),
\label{5.8}
\end{equation}
where entering functions defined in Eqs.(\ref{4.20}) and
(\ref{4.22}).

In the case of coaxial beams ${\bf x}_0=0,~J=0$ one can take the
integral in (\ref{5.8}) over one of variables (for definiteness
over $s_2$) using the formula
\begin{equation}
\int_{0}^{\infty}\frac{dx}{(a_z+b_z x)^{3/2}(a_y+b_y x)^{1/2}}=
\frac{2}{a_z\sqrt{b_z b_y}+b_z\sqrt{a_z a_y}}.
\label{5.8a}
\end{equation}
After this we have the simple integral over $s \equiv s_1$
\begin{eqnarray}
&& J_{-}(0)=\sqrt{1+\delta_z}\sqrt{1+\delta_y}(J_z+J_y),
\nonumber \\
&& J_{z,y}=\int_{0}^{\infty}D_{z,y}(s)ds,\quad
D_{z,y}=\frac{1}{a_{z,y}\sqrt{b_z b_y}+b_{z,y}\sqrt{a_z a_y}},
 \label{5.8b}
\end{eqnarray}
where
\begin{equation}
a_{z,y}=s(1+\delta_{z,y})+2\sigma_{z,y}^2 (2+\delta_{z,y}),\quad
b_{z,y}=\frac{s}{2\Delta_{z,y}^2}+1+\delta_{z,y},\quad
\delta_{z,y}= \frac{\sigma_{z,y}^2}{\Delta_{z,y}^2},
\label{5.8c}
\end{equation}

The cross section (\ref{5.7}) differs from Eq.(24) of \cite{BKS1}
because the subtraction term $J_{-}$ is included. Without this
term, generally speaking, the bremsstrahlung cross section would
be qualitatively erroneous. In particular an appearance of the
term $J_{-}$ violates, generally speaking, the symmetry of
radiation cross section in opposite directions in
$e^{-}e^{-}~(e^{-}e^{+})$ collisions.

To elucidate the qualitative features of narrow beams
bremsstrahlung process we consider the case of round beams where
the calculation becomes more simple:
\begin{eqnarray}
&& \sigma_z=\sigma_y=\sigma,\quad \Delta_z=\Delta_y=\Delta, \quad
\Sigma_z^2=\Sigma_y^2=\Sigma^2=\frac{1}{2(\sigma^2+\Delta^2)},
\nonumber \\
&& b=a,\quad b_{1,2}=a_{1,2},\quad B=A, \quad
\delta=\frac{\sigma^2}{\Delta^2}.
\label{5.9}
\end{eqnarray}
We consider first the case of coaxial beams (${\bf x}_0=0,~J=0$),
\begin{eqnarray}
&& J_{-}=(1+\delta)\int_{0}^{\infty}
\frac{ds}{[s(1+\delta)+2+\delta][s\delta +1+\delta]}
\nonumber \\
&& =(1+\delta) \ln \frac{(1+\delta)^2}{\delta(2+\delta)}.
\label{5.10}
\end{eqnarray}
In the limiting cases the function $J_{-}$ has the form
\begin{equation}
 J_{-}(\delta \gg 1) \simeq \frac{1}{\delta},\quad
 J_{-}(\delta=1) = 2\ln\frac{4}{3}, \quad J_{-}(\delta \ll 1)
 \simeq \ln \frac{1}{2\delta}.
 \label{5.11}
\end{equation}
In the first case the subtraction term $J_{-}$ is small. For the
beams of the same size the subtraction term $J_{-}$ contributes
to the constant entering into the expression for the cross
section. The subtraction term $J_{-}$ modifies essentially the
cross section in the case when the radius of target beam is much
smaller than the radius of radiating beam. In this case the cross
section (\ref{5.7}) contains the combination
\begin{equation}
 \ln \frac{m^2}{4\Sigma^2}-J_{-} \simeq \ln \frac{m^2\Delta^2}{2}
 -\ln \frac{\Delta^2}{2\sigma^2}=\ln (m\sigma)^2.
\label{5.12}
\end{equation}
So in the all cases considered above the cross section defines
the transverse dimension of target beam.

When the axes of round beams are displaced with respect each other
in the transverse plane the integral in (\ref{5.4}) is
\begin{eqnarray}
&&J=\int_{0}^{\infty}
\left[\exp\left(\frac{d}{x+1}\right)-1\right]\frac{dx}{x+1}={\rm
Ei}(d)-C-\ln d ,
\nonumber \\
&& d={\bf x}_0^2\Sigma^2 =
\frac{x_0^2+y_0^2}{2(\Delta^2+\sigma^2)}.
\label{5.13}
\end{eqnarray}

It is convenient in this case to calculate the function $J_{-}$
using straightforwardly Eq.(\ref{4.18}) where we omit the term
with $\eta^2$ in the exponent of integrand
\begin{equation}
{\bf I}_{cr}= \mbox{\boldmath$\varrho$}\int_{0}^{\infty}
\exp\left(-\frac{\mbox{\boldmath$\varrho$}^2}{s+2\sigma^2}\right)
\frac{ds}{(s+2\sigma^2)^2}=
\frac{\mbox{\boldmath$\varrho$}}{\mbox{\boldmath$\varrho$}^2}\left[1-
\exp\left(-\frac{\mbox{\boldmath$\varrho$}^2}{2\sigma^2}\right)\right]
\label{5.14}
\end{equation}
Substituting this expression ({\bf I} is defined in
Eq.(\ref{4.15})) into the subtraction term Eq.(\ref{3.12}) and
using the exponential parametrization
\[
\frac{1}{\mbox{\boldmath$\varrho$}^2}=\int_{0}^{\infty}
\exp(-\mbox{\boldmath$\varrho$}^2s)ds
\]
we obtain
\begin{eqnarray}
&& J_{-}=\frac{a e^d}{\pi \Sigma^2} \int_{0}^{\infty}ds \int
d^2\varrho \exp(-\mbox{\boldmath$\varrho$}^2s) \left[1-
\exp\left(-\frac{\mbox{\boldmath$\varrho$}^2}{2\sigma^2}\right)\right]
\exp(-a(\mbox{\boldmath$\varrho$}+{\bf x}_0)^2)
\nonumber \\
&& = \frac{a e^{d-d_1}}{\Sigma^2}
\int_{0}^{\infty}\Bigg[\frac{1}{s+a}\exp\left( \frac{d_1  a
}{s+a}\right)-2\frac{1}{s+a+\sigma^{-2}/2}\exp\left(
d_1\frac{a}{s+a+\sigma^{-2}/2} \right)
\nonumber \\
&& + \frac{1}{s+a+\sigma^{-2}}\exp\left(
d_1\frac{a}{s+a+\sigma^{-2}} \right)\Bigg]ds= \frac{a
e^{d-d_1}}{\Sigma^2}\Bigg[ {\rm Ei}(d_1)-2{\rm Ei}\left(d_1
\frac{\sigma^2}{\sigma^2+\Delta^2} \right)
\nonumber \\
&& +{\rm Ei}\left(d_1 \frac{\sigma^2}{\sigma^2+2\Delta^2}
\right)\Bigg],\quad d_1=a{\bf
x}_0^2=\frac{z_0^2+y_0^2}{2\Delta^2}.
\label{5.15}
\end{eqnarray}
In the limit $d_1 \rightarrow 0$ the last expression goes over to
Eq.(\ref{5.10}).

When the displacement of the axes of colliding beams is large
enough $({\bf x}_0^2 \gg \sigma^2+\Delta^2)$ one use the
asymptotic expansion of the function Ei($z$) in (\ref{5.15}):
\begin{equation}
{\rm Ei}(z) \simeq \frac{e^z}{z}\left( 1+\frac{1}{z}\right),\quad
z \gg 1.
\label{5.16}
\end{equation}
In this case the main terms in the difference $J-J_{-}$ in
Eq.(\ref{5.7}) are canceled:
\begin{equation}
J-J_{-} \simeq \frac{e^{d}}{d}\left(
\frac{1}{d}-\frac{1}{d_1}\right) = \frac{2 e^{d}}{d}
\frac{\sigma^2}{{\bf x}_0^2}.
\label{5.16a}
\end{equation}
The compensation of the main terms in (\ref{5.16}) is due to the
fact that the incoherent scattering originates on the
fluctuations of the potential of the target (scattering) beam.
Correspondingly we have for the mean square of the momentum
transfer dispersion at the large distance from the target beam
\begin{eqnarray}
&& <{\bf q}^2(\mbox{\boldmath$\varrho$})>-<{\bf
q}(\mbox{\boldmath$\varrho$})>^2 \propto \left< \frac{1}{({\bf
x}_0+\mbox{\boldmath$\varrho$})^2}-\frac{1}{{\bf x}_0^2}\right>
\nonumber \\
&& \simeq \left< \frac{4({\bf x}_0
\mbox{\boldmath$\varrho$})^2}{{\bf
x}_0^6}-\frac{\mbox{\boldmath$\varrho$}^2}{{\bf x}_0^4}\right> =
\frac{<\mbox{\boldmath$\varrho$}^2>}{{\bf
x}_0^4}=\frac{2\sigma^2}{{\bf x}_0^4}.
\label{5.17}
\end{eqnarray}
Substituting (\ref{5.16a}) into Eq.(\ref{5.7}) and multiplying the
result by the luminosity (\ref{4.3}):
\begin{equation}
{\cal L}=N_c N_r\frac{\Sigma^2}{\pi}\exp(-{\bf x}_0^2 \Sigma^2)
 \label{5.18}
\end{equation}
we have for the probability of bremsstrahlung of round beams
moving apart at large distance
\begin{eqnarray}
&&dw_{\gamma} \simeq 4N_c
N_r\frac{\alpha^3}{\pi}\lambda_c^2\Sigma^2
\frac{\varepsilon'}{\varepsilon}\frac{d\omega}{\omega}\left(
v-\frac{2}{3} \right)
\nonumber \\
&& \times\left[ \exp(-{\bf x}_0^2 \Sigma^2) \ln \frac{m}{\Sigma}+
\frac{\sigma^2\Sigma^2}{({\bf x}_0^2\Sigma^2)^2}+O(\exp(-{\bf
x}_0^2 \Sigma^2))\right]
\nonumber \\
&& \Sigma^2=\frac{1}{2(\Delta^2+\sigma^2)},\quad  {\bf x}_0^2
\Sigma^2=\frac{z_0^2+y_0^2}{2(\Delta^2+\sigma^2)} \gg 1, \quad
q_{min}^2(z_0^2+y_0^2) \ll 1.
\label{5.19}
\end{eqnarray}
According to (\ref{5.19}) when ${\bf x}_0^2$ increases so that
one can neglect the first term in square brackets, the probability
of bremsstrahlung of the round beams diminishes as a power of
distance between beams ($\propto \sigma^2/{\bf x}_0^4$). The
cross section Eq.(\ref{5.7}) in this case grows exponentially as
$e^d/d^2$. Let us note that without the subtraction term one has
erroneous qualitative behaviour of probability ($\propto 1/{\bf
x}_0^2$).  These circumstances explain also Eq.(\ref{5.12}) for
the coaxial beams: at integration over $d^2\varrho$ the region
contributes where $<{\bf q}^2(\mbox{\boldmath$\varrho$})>- <{\bf
q}(\mbox{\boldmath$\varrho$})>^2 \propto
1/\mbox{\boldmath$\varrho$}^2$, so that $\varrho \leq \sigma$.

Let us consider now the general case $\Sigma_z \neq \Sigma_y$ for
enough large displacement of beams ${\bf x}_0^2 \gg
\Sigma_{z,y}^{-2}$. In this case the main contribution into the
integral ${\bf I}({\bf x})$ (for $\eta^2=0$) in
Eqs.(\ref{4.15}),(\ref{4.18}) at large $|{\bf x}| \simeq |{\bf
x}_0|$ (see Eq.(\ref{3.12})) are given by large values $s \sim
{\bf x}_0^2 \gg \sigma_{z,y}^2$. Expanding the integrand over the
powers $\sigma_{z,y}^2/s$ and keeping after integration the two
main terms of the decomposition over $1/{\bf x}^2$ we get
\begin{equation}
{\bf I}^2({\bf x}) \simeq \frac{1}{{\bf
x}^2}\left[1+\frac{2}{({\bf x}^2)^2} (y^2-z^2)(\sigma_y^2 -
\sigma_z^2)\right].
 \label{5.20}
\end{equation}
Expanding the function $1/({\bf x}_0+\mbox{\boldmath$\xi$})^2$
over the powers $\xi/x_0$ at the integration over
$\mbox{\boldmath$\xi$}={\bf x}-{\bf x}_0$ in Eq.(\ref{3.12})) we
find
\begin{eqnarray}
&&\int {\bf I}^2({\bf
x}_0+\mbox{\boldmath$\xi$})w_r(\mbox{\boldmath$\xi$})d^2\xi
\simeq \frac{1}{{\bf x}_0^2} \Bigg[1 + \frac{4}{({\bf
x}_0^2)^2}(z_0^2\Delta_z^2+y_0^2\Delta_y^2)
\nonumber \\
&& -\frac{\mbox{\boldmath$\Delta$}^2}{{\bf x}_0^2}+ \frac{2}{({\bf
x}_0^2)^2}(y_0^2-z_0^2)(\sigma_y^2 - \sigma_z^2)\Bigg], \quad
\mbox{\boldmath$\Delta$}^2=\Delta_z^2+\Delta_y^2.
\label{5.21}
\end{eqnarray}
In this case the region $t \sim 1/{\bf x}_0^2 \ll \Sigma_{z,y}^2$
contributes into the integral $J$ Eq.(\ref{5.4})). Expanding the
integrand over the powers $t\Sigma_{z,y}^{-2}$ and keeping the
two main terms of decomposition over $1/{\bf x}_0^2$ we have
\begin{eqnarray}
&& J \simeq \frac{1}{\Sigma_z\Sigma_y {\bf x}_0^2}
\exp(z_0^2\Sigma_z^2+y_0^2\Sigma_y^2)
\Bigg\{1-\frac{\mbox{\boldmath$\sigma$}^2+\mbox{\boldmath$\Delta$}^2}{{\bf
x}_0^2}
\nonumber \\
&& +\frac{4}{({\bf x}_0^2)^2}\left[ z_0^2(\sigma_z^2+\Delta_z^2)
+y_0^2 (\sigma_y^2+\Delta_y^2)\right]\Bigg\}, \quad
\mbox{\boldmath$\sigma$}^2 =\sigma_z^2+\sigma_y^2.
\label{5.22}
\end{eqnarray}
For the difference $J-J_{-}$ we obtain finally
\begin{equation}
J-J_{-} =
\frac{1}{\Sigma_z\Sigma_y}\exp(z_0^2\Sigma_z^2+y_0^2\Sigma_y^2)
\frac{\mbox{\boldmath$\sigma$}^2}{({\bf x}_0^2)^2}.
 \label{5.23}
\end{equation}

\section{Narrow flat beams ($\sigma_z \ll \sigma_y, \Delta_z \ll \Delta_y)$}

\setcounter{equation}{0}

Let us begin with the coaxial beams. We consider first the case
where the size of radiating beam is much larger than size of
target beam ($\delta_{z,y} \ll 1$). In this case one can neglect
the terms $\propto \delta_{z,y}, \sigma_z^2, \Delta_y^{-2}$ in the
functions $a_{z,y}$ and $b_{z,y}$ in the integral in
Eq.(\ref{5.8b}). Within this accuracy
\begin{equation}
a_z \simeq s,\quad a_y \simeq s +4\sigma_y^2,\quad b_z  \simeq
\frac{s}{2\Delta_z^2}+1, \quad b_y  \simeq 1.
 \label{6.1}
\end{equation}
After substitution in the integral $J_y$ in Eq.(\ref{5.8b}) $s
\rightarrow 4\sigma_y^2 s$ one gets
\begin{equation}
J_y(\kappa) =  \int_{0}^{\infty}
\frac{ds}{\sqrt{s+1}(\sqrt{s}+\sqrt{s+1}\sqrt{1+2\kappa
s})},\quad \kappa=\frac{\sigma_y^2}{\Delta_z^2}.
 \label{6.2}
\end{equation}
After substitution in the integral $J_z$ in Eq.(\ref{5.8b}) $s
\rightarrow 2\Delta_z^2/s$ one gets $J_z=J_y$ so that
\begin{eqnarray}
&& J_{-}(\kappa)=2\sqrt{1+\delta_z}\sqrt{1+\delta_y}J_y(\kappa)
\simeq 2J_y(\kappa),
\nonumber \\
&&  J_{-}(\kappa \ll 1) \simeq
\ln \frac{8}{\kappa}, \quad J_{-}(\kappa \gg 1) \simeq \pi
\sqrt{\frac{2}{\kappa}}.
\label{6.3}
\end{eqnarray}
It is seen from the last equation that at $\Delta_z \ll \sigma_y$
the contribution of the term $J_{-}$ into the cross section
Eq.(\ref{5.7}) is relatively small. In the opposite case
$\Delta_z \gg \sigma_y$ this contribution leads to change of the
logarithm argument in Eq.(\ref{5.7})
\begin{equation}
2 \ln \frac{m}{(\Sigma_z+\Sigma_y)} - \ln \frac{8}{\kappa} \simeq
2(\ln (\sqrt{2}m \Delta_z)-\ln (2\sqrt{2}
\frac{\Delta_z}{\sigma_y}))=2 \ln \frac{m\sigma_y}{2}.
 \label{6.4}
\end{equation}
This is a new qualitative result.

In the opposite case when the size of radiating beam is smaller
or is of the order of size of target beam ($\delta_{z,y} \geq 1$)
the contribution into the integral $J_z$ in Eq.(\ref{5.8b}) gives
the region $s \sim \sigma_z^2$ and into the integral $J_y$ the
region $s \sim \sigma_y^2$. Performing in the integral $J_z$ the
substitution $s \rightarrow 2\sigma_z^2 s$ and in the integral
$J_y$ the substitution $s \rightarrow 2\sigma_y^2/s$ one gets
\begin{eqnarray}
&& J_z \simeq \frac{\sigma_z}{\sqrt{2+\delta_y}\sigma_y}
\int_{0}^{\infty}\frac{ds}{((s+1)\delta_z+1)\sqrt{s(1+\delta_z)+2+\delta_z}}
\nonumber \\
&& = \frac{2}{\sqrt{2+\delta_y}}\frac{\Delta_z}{\sigma_y}\arctan
\frac{1}{\sqrt{\delta_z(2+\delta_z)}};
\nonumber \\
&& J_y \simeq \frac{\Delta_z}{\sigma_y}
\int_{0}^{\infty}\frac{ds}{((s+1)(\delta_y+1)+s)\sqrt{(s+1)\delta_y+s}}
\nonumber \\
&& = \frac{2}{\sqrt{2+\delta_y}}\frac{\Delta_z}{\sigma_y}\arctan
\frac{1}{\sqrt{\delta_y(2+\delta_y)}};
\nonumber \\
&& J_{-}=\sqrt{1+\delta_z}\sqrt{1+\delta_y}(J_z+J_y)
\nonumber \\
&& =
\frac{2\sqrt{1+\delta_z}\sqrt{1+\delta_y}}{\sqrt{2+\delta_y}}
\frac{\Delta_z}{\sigma_y}
\left(\arctan \frac{1}{\sqrt{\delta_z(2+\delta_z)}}+\arctan
\frac{1}{\sqrt{\delta_y(2+\delta_y)}} \right).
 \label{6.5}
\end{eqnarray}
In the case $\delta_{z,y} \ll 1,~ \Delta_z \ll \sigma_y$ this
formula is consistent with Eq.(\ref{6.3}).

Now we go over to the case of the displaced beams. For enough
large displacement of the beams the formulas (\ref{5.7}) and
(\ref{5.23}) are valid. So the intermediate case is of interest.
As an example we consider the case $\sigma_y^2 \gg z_0^2 \gg
\sigma_z^2+\Delta_z^2,~y_0^2 \ll \sigma_y^2$. In this case the
contribution to the integral in (\ref{5.4}) gives the interval
$\Sigma_y^2 \ll t \sim z_0^{-2} \ll \Sigma_z^2$. Keeping the main
terms of decomposition over $t\Sigma_z^{-2}\ll 1$ and
$t\Sigma_y^{-2}\gg 1$ we have
\begin{equation}
J \simeq \frac{1}{\Sigma_z} \int_{0}^{\infty}\exp \left( z_0^2
\Sigma_z^2-z_0^2 t
\right)\frac{dt}{\sqrt{t}}=\frac{\sqrt{\pi}}{z_0
\Sigma_z}\exp(z_0^2 \Sigma_z^2).
 \label{6.6}
\end{equation}
Under these conditions (${\bf x}_0^2 \ll \sigma_y^2$) the
contribution into the integral for $J_{-}$ in (\ref{5.8})of the
term in the function $G(s_1, s_2, {\bf x}_0 )$ Eq.(\ref{4.20}) in
the square brackets containing $b_1b_2/B$ is defined by the
function $J_y$ in Eq.(\ref{6.5}) to within the terms $\sim
z_0/\sigma_y$. In the term containing $a_1a_2/A$ (which we denote
by $J_{-}^{(z)}$) the main contribution gives the summand
$z_0^2a^2/A^2$ in the interval $\sigma_y^2 \gg s_{1,2} \sim z_0^2
\gg \sigma_z^2$ where
\begin{equation}
a_{1,2} \simeq \frac{1}{s_{1,2}},\quad b_{1,2} \simeq
\frac{1}{2\sigma_y^2},\quad A \simeq a, \quad B \simeq
\frac{1}{\sigma_y^2}+\frac{1}{2\Delta_y^2}.
 \label{6.7}
\end{equation}
As a result we obtain
\begin{eqnarray}
&& J_{-}^{(z)}({\bf x}_0) \simeq
\frac{z_0^2}{2\Sigma_z\Sigma_y\sigma_y^2}\sqrt{\frac{b}{B}}
e^{d_z} \int_{0}^{\infty}\frac{ds_1}{s_1^{3/2}}
\int_{0}^{\infty}\exp\left(-z_0^2\left(
\frac{1}{s_1}+\frac{1}{s_2} \right)\right)\frac{ds_2}{s_2^{3/2}}
\nonumber \\
&& =\pi \frac{\Delta_z}{\sigma_y}
\frac{\sqrt{1+\delta_z}\sqrt{1+\delta_y}}{\sqrt{2+\delta_y}}e^{d_z},
\nonumber \\
&& J-J_{-} \simeq \sqrt{\frac{\pi}{d_z}}e^{d_z}h(z_0),\quad
d_z=z_0^2\Sigma_z^2 ,
\nonumber \\
&& h(z_0)=1-\frac{\sqrt{\pi(1+\delta_y)}}{\sqrt{2(2+\delta_y)}}
\frac{z_0}{\sigma_y}\left(
1+\frac{2}{\pi}\arctan\frac{1}{\sqrt{\delta_y(2+\delta_y)}}\right).
 \label{6.8}
\end{eqnarray}
It should be noted that for the flat beams the probability of
radiation as a function of distance between beams (for the
considered interval) decreases more slowly $\propto 1/\sqrt{d_z}$
than for the round beams given in Eq.(\ref{5.19})
\begin{equation}
dw_{\gamma}^{fl}\simeq
4N_cN_r\frac{\alpha^3}{\pi}\lambda_c^2\Sigma_z\Sigma_y\frac{\varepsilon'}{\varepsilon}
\frac{d\omega}{\omega}\left( v-\frac{2}{3}\right)\left[
e^{-d_z}\ln
\frac{m}{\Sigma_z}+\frac{1}{2}\sqrt{\frac{\pi}{d_z}}h(z_0)\right].
 \label{6.9}
\end{equation}
Compensation in the difference $J-J_{-}$ begins in the region
$z_0 \sim \sigma_y+\Delta_y$ were Eq.(\ref{6.8}) is not valid and
one have to use more accurate Eq.(\ref{5.8}). In the region $z_0
\gg \sigma_y+\Delta_y$ the probability of radiation decreases as
$1/z_0^4$ according to  Eqs.(\ref{4.4}), (\ref{5.7}), (\ref{5.23})
provided that one can neglect the exponential term in the square
brackets in Eq.(\ref{6.9}) (compare with Eq.(\ref{5.19}))
\begin{equation}
dw_{\gamma}^{fl}(z_0) \simeq
2N_cN_r\frac{\alpha^3}{\pi}\frac{\lambda_c^2
\sigma_y^2}{z_0^4}\frac{\varepsilon'}{\varepsilon} \left(
v-\frac{2}{3}\right)\frac{d\omega}{\omega}, \quad z_0 \gg y_0.
 \label{6.10}
\end{equation}

\section{Observation of beam-size effect}

\setcounter{equation}{0}

Above we calculated the incoherent bremsstrahlung spectrum at
collision of electron and positron beams with finite transverse
dimensions. This spectrum differs from spectrum found previously
in \cite{BKS1}, \cite{BD}, \cite{KPS} because here (in contrast
to previous papers) we subtract the coherent contribution. In
general expression for correction to the probability of photon
emission (\ref{3.11}) the subtraction term is $F^{(2)}(\omega,
\zeta)$. For the coaxial beams for numerical calculation it is
convenient to use Eqs.(\ref{4.9}), (\ref{4.19}) and (\ref{4.20}).
In the last equation one have to put $y_0=z_0=0$. In the case of
collision of narrow beams the subtraction term in the
bremsstrahlung spectrum (\ref{5.7}) is $J_{-}$. The dimensions of
beams in the experiment \cite{exp1} were
$\sigma_z=\Delta_z=24~\mu m, \sigma_y=\Delta_y=450~\mu m$, so
this is the case of flat beams. The estimate for this case
(\ref{6.5}) gives $J_{-} \simeq (4/3\sqrt{3})\pi \sigma_z/\sigma_y
\ll 1$. This term is much smaller than other terms in (\ref{5.7}).
This means that for this case the correction to the spectrum
calculated in \cite{BKS1} is very small.

The result of calculation and VEPP4 (INP, Novosibirsk) data are
presented in Fig.1 where the bremsstrahlung intensity spectrum
$\omega d\sigma/d\omega$ is given in units $2\alpha r_0^2$ versus
the photon energy in units of initial electron energy
($x=\omega/\varepsilon$). The upper curve is the standard QED
spectrum, the three close curves below are calculated using
Eqs.(\ref{4.9}) and (\ref{4.19}) for the different vertical
dimensions of colliding beams (equal for two colliding beams
$\sigma=\sigma_z=\Delta_z$):$\sigma=20~\mu m$ (bottom),
$\sigma=24~\mu m$ (middle), $\sigma=27~\mu m$ (top) (this is just
the 1-sigma dispersion for the beams used in the experiment). We
want to emphasize that all the theoretical curves are calculated
to within the relativistic accuracy (the discarded terms are of
the order $m/\varepsilon$). It is seen that the effect of the
small transverse dimensions is very essential in soft part of
spectrum (at $\omega/\varepsilon = 10^{-4}$ the spectral curve
diminishes in 25 \%), while for $\omega/\varepsilon > 10^{-1}$
the effect becomes negligible. The data measured in \cite{exp1}
are presented as circles (experiment in 1980) and as triangles
(experiment in 1981) with 6 \% systematic error as obtained in
\cite{exp1} (while the statistical errors are negligible). This
presentation is somewhat different from \cite{exp1}. It is seen
that the data points are situated systematically below the theory
curves but the difference is not exceed the 2-sigma level
\cite{exp1}. It should be noted that this is true also in the
hard part of spectrum where the beam-size effect is very small.

The last remark is connected with the radiative corrections (RC).
The RC to the spectrum of double bremsstrahlung \cite{BG} (this
was the normalization process) are essential (of the order 10 \%)
and were taken into account. The RC to the bremsstrahlung spectrum
\cite{KL} are very small (less than 0.4 \%) and may be neglected.
It should be noted that the RC to the bremsstrahlung spectrum are
insensitive to the effect of small transverse dimensions.

The dependence of bremsstrahlung spectrum on beams
characteristics was measures specifically in \cite{exp1}. The
first is the dependence of bremsstrahlung spectrum on vertical
sizes of beams $\sigma_z$. It is calculated using Eqs.(\ref{4.9})
and (\ref{4.19}) for $\omega/\varepsilon = 10^{-3}$. The result is
shown in units $2\alpha r_0^2$ in Fig.2. The data is taken from
Fig.7 in \cite{exp1}. The second is the measurement of dependence
of bremsstrahlung spectrum on the vertical displacement of beams
$z_0$. It is calculated using Eqs.(\ref{5.4}) and (\ref{5.8}) for
$\omega/\varepsilon = 10^{-3}$. Because of displacement it is
necessary to normalize the spectrum on the luminosity
\[
{\cal L}=N_c N_r\frac{\Sigma_z
\Sigma_y}{\pi}\exp(-z_0^2\Sigma_z^2),
\]
see Eq.(\ref{4.3}). This means that when we compare the
bremsstrahlung process (where the beam-size effect is essential)
with some other process like double bremsstrahlung used in
\cite{exp1} (which is insensitive to the effect) we have to
multiply the cross section of the last process by the luminosity
${\cal L}$. This is seen in estimate Eq.(\ref{6.9}): after taking
out the exponent $e^{-d_z}$ we have the luminosity as the
external factor and in expression for ratio
$N_{\gamma}/N_{2\gamma}$ (which was observed in \cite{exp1}) the
cross section of double bremsstrahlung will be multiplied by the
luminosity. After this operation the second term in square
brackets will contain the combination $e^{d_z}h(z_0)/\sqrt{d_z}$
which grows exponentially with the displacement $z_0$ increase.
The normalized bremsstrahlung spectrum is shown in units $2\alpha
r_0^2$ in Fig.3. So, the very fast (exponential) increase with
$z_0$ is due to fast decrease with $z_0$ of the double
bremsstrahlung probability for the displaced beams. The data is
taken from Fig.8 in \cite{exp1}. It should be noted that in soft
part of spectrum the dependence on photon energy $\omega$ is very
weak. It is seen in these figures that there is quite reasonable
agreement between theory and data just as in \cite{exp1}. This
means that contribution of $J_{-}$ term which is calculated only
in the present paper is relatively small.

One more measurement of beam-size effect was performed at HERA
electron-proton collider (DESY, Germany) \cite{P}. The electron
beam energy was $\varepsilon$=27.5~GeV, the proton beam energy was
$\varepsilon_p$=820~GeV. The standard bremsstrahlung spectrum for
this case is given by Eq.(\ref{5.6}) where $q_{min}$ should be
substituted:
\begin{equation}
q_{min} \rightarrow q_{min}^D=\frac{\omega m^2 m_p}{4\varepsilon_p
\varepsilon \varepsilon'}, \label{7.1}
\end{equation}
here $m_p$ is the proton mass. In this situation the
characteristic length is $l_{f0}^D=1/q_{min}^D$ and at the photon
energy $\omega=1$~GeV one has $l_{f0}^D \sim 2$~mm. Since the
beam sizes at HERA are much smaller than this characteristic
length, the beam-size effect can be observed at HERA. The
parameters of beam in this experiment were (in our notation):
$\sigma_z=\Delta_z=(50 \div 58) \mu m$, $\sigma_y=\Delta_y=(250
\div 290) \mu m$. In part of runs the displaced beams were used
with $z_0=20~\mu m$ and $y_0=100~\mu m$. The bremsstrahlung
intensity spectrum $\omega d\sigma/d\omega$ in units $2\alpha
r_0^2$ versus the photon energy in the units of initial electron
energy ($x=\omega/\varepsilon$) for the HERA experiment is given
in Fig.4. The upper curve is the standard QED spectrum. We
calculated the spectrum with beam-size effect taken into account
for three sets of beams parameters; the set 1:
$\sigma_z=\Delta_z=50~\mu m, \sigma_y=\Delta_y=250~\mu m,
z_0=y_0=0$, the set 2: $\sigma_z=\Delta_z=50~\mu m,
\sigma_y=\Delta_y=250~\mu m, z_0=20 ~\mu m, y_0=0$, the set 3:
$\sigma_z=\Delta_z=54~\mu m, \sigma_y=\Delta_y=250~\mu m,
z_0=y_0=0$. The result of calculation is seen as two close curves
below, the top curve is for the set 3, while the bottom curve is
actually two merged curves for the sets 1 and 2. Since the ratio
of the vertical and the horizontal dimensions is not very small,
the general formulas were used in calculation: for coaxial beams
Eqs. (\ref{4.10}) and (\ref{4.19}),and for displaced beams Eqs.
(\ref{4.13}) and (\ref{4.19}). It should be noted that the
contribution of subtraction term (Eq.(\ref{4.19})) is quite
essential (more than 10\%) for the beam parameters used at HERA.
The data are taken from Fig.5c in \cite{P}. The errors are the
recalculated overall systematic error given in \cite{P}. It is
seen that there is a quite satisfactory agreement of theory and
data. The final data are given in \cite{P} also as the averaged
relative difference
$\delta=(d\sigma_{QED}-d\sigma_{bs})/d\sigma_{QED}$ (where
$d\sigma_{QED}$ is the standard QED spectrum, $\sigma_{bs}$ is
the result of calculation with the beam-size effect taken into
account) over the whole interval of photon energies (2-8~GeV),
e.g. for the set 1 $\delta_{ex}=(3.28 \pm 0.7)\%$, for the set 2
$\delta_{ex}=(3.57 \pm 0.7)\%$, for the set 3 $\delta_{ex}=(3.06
\pm 0.7)\%$,  \cite{P}. The averaged $<\delta>$ over the interval
$0.07 \leq x \leq 0.28$ (or 1.95~GeV$ \leq \omega \leq$ 7.7~GeV)
in our calculation are for the set 1 is $<\delta>$=2.69\%, for
the set 2 is $<\delta>$=2.65\%, for the set 3 is
$<\delta>$=2.54\%. So, for these data there is also a
satisfactory agreement of data and theory (at the 1-sigma level,
except set 2 where the difference is slightly larger).

So, the beam-size effect discovered at BINP (Novosibirsk) was
confirmed at DESY (Germany). Of course, more accurate measurement
is desirable to verify that we entirely understand this mechanism
of deviation from standard QED.

\section{Conclusion}
\setcounter{equation}{0}

Above the influence of the finite transverse size of the colliding
beams on the incoherent bremsstrahlung process is investigated.
Previously (see papers \cite{BKS1}, \cite{BD}, \cite{KPS},
\cite{KSS}) for analysis of this effect an incomplete expression
for the bremsstrahlung intensity spectrum was used because in it
the subtraction was not fulfilled. It is necessary to carry out
this subtraction for the extraction of pure fluctuation process
which is just the incoherent bremsstrahlung. We implement this
procedure in the present paper. We indicated the cases where the
results without the subtraction term are qualitatively erroneous.
The first this is the case when the transverse sizes of scattering
beam are much smaller than the corresponding sizes of radiating
beam. For coaxial round beams see e.g. Eq.(\ref{5.12}) and for
flat beams Eq.(\ref{6.4}). In contrast to previous papers here we
draw a conclusion that the bremsstrahlung cross section is
determined by the transverse sizes of scattering beam.

The new qualitative result is deduced for the case when the
displacement of beams is enough large. Then the square of
momentum transfer dispersion, which determines the bremsstrahlung
cross section, decreases with displacement increase faster than
mean value the momentum transfer squared (see Eqs.(\ref{5.17}),
(\ref{5.23})). As it was noted in Sec.7, it is necessary to
normalize the spectrum on the luminosity for displaced beams.
Then the bremsstrahlung cross section grows exponentially with
displacement increase. This very fast (exponential) increase with
$z_0$ is due to fast decrease with $z_0$ normalization process
probability for displaced beams.

For Gaussian beams the expression for the bremsstrahlung spectrum
is obtained in the form of double integrals convenient for
numerical calculations (see Eqs.(\ref{4.9}), (\ref{4.19}) and
(\ref{4.20})).For soft part of spectrum we deduced the general
expression for spectrum which is independent of minimal momentum
transfer $q_{min}$ and is defined only by transverse size of
beams (see Eqs.(\ref{5.3}), (\ref{5.4}) and
(\ref{5.7})-(\ref{5.8c})).

The important feature of the considered beam-size effect is smooth
decrease of radiation probability with growth of displacement of
beams. For the flat beams we see in Eqs. (\ref{6.9}), (\ref{6.10})
that the main (logarithmic) term in expression for the probability
decreases exponentially ($\propto \exp(-z_0^2\Sigma_z^2)$ as
luminosity), but there is the specific long-range term $\propto
1/z_0$ which results in quite appreciable radiation probability
even in the case when beam the displacement is large. This
phenomenon may be helpful for tuning of high-energy
electron-positron colliders. As an example we consider the
"typical" collider were the beam energy is $\varepsilon=500~$GeV
and the beam dimensions are equal and $\sigma_z=5 nm$ and
$\sigma_y=100 nm$. The beam-size effect in this collider is very
strong and for $x=10^{-3}$ the intensity spectrum is only $\sim
0.3$ part of the standard $\omega d\sigma_{QED}(\omega)/d\omega$.
The dependence of bremsstrahlung probability on the displacement
distance $z_0$ (in $nm$) is shown in Fig.5. It is calculated
using Eqs. (\ref{5.6})- \ref{5.8}) for soft photons with
$x=10^{-3}$ (the asymptotic formulas (\ref{6.9}) are (\ref{6.10})
are not enough accurate in this case). Actually the dependence on
photon energy is contained in the external factor
$(1-x)(v(x)-2/3)$. The curve in Fig.5 reflects the main features
mentioned above. One can see that even for $z_0$=100
($z_0=20~\sigma_z$) the cross section is $\sim 0.002$ part of
very large bremsstrahlung probability at head-on collision of
beams. So, measuring the radiation for displaced beams one can
estimate magnitude of displacement of beams. This information may
be useful for beam tuning.

{\bf Acknowledgments}
%\vspace{0.2cm}

We would like to thank Prof. Krzysztof Piotrzkowsky for additional
information about the HERA experiment. This work was supported in
part by the Russian Fund of Basic Research under Grant
00-02-18007.

\newpage

{\bf Figure captions}

\vspace{15mm}
\begin{itemize}

\item {\bf Fig.1} The bremsstrahlung intensity spectrum $\omega
d\sigma/d\omega$ in units $2\alpha r_0^2$ versus the photon
energy in units of initial electron energy
($x=\omega/\varepsilon$) for VEPP4 experiment. The upper curve is
the standard QED spectrum, the three close curves below are
calculated  for the different vertical dimensions of colliding
beams (equal for two colliding beams
$\sigma=\sigma_z=\Delta_z$):$\sigma=20~\mu m$ (bottom),
$\sigma=24~\mu m$ (middle), $\sigma=27~\mu m$ (top). The data
measured in \cite{exp1} are presented as circles (the experiment
in 1980) and as triangles (the experiment in 1981) with 6 \%
systematic error as obtained in \cite{exp1}.

\item {\bf Fig.2} The bremsstrahlung intensity spectrum $\omega
d\sigma/d\omega$ in units $2\alpha r_0^2$ versus the vertical
sizes of beams $\sigma_z$ (in $\mu m$). The data taken from
\cite{exp1}.

\item {\bf Fig.3} The normalized to luminosity ${\cal L}$ the
bremsstrahlung intensity spectrum $\omega d\sigma/d\omega$ in
units $2\alpha r_0^2$ versus the vertical displacement of beams
$z_0$ (in $\mu m$). The data taken from \cite{exp1}.

\item {\bf Fig.4} The bremsstrahlung intensity spectrum
$\omega d\sigma/d\omega$ in units $2\alpha r_0^2$ versus the
photon energy in units of initial electron energy
($x=\omega/\varepsilon$) for the HERA experiment. The upper curve
is the standard QED spectrum, the two close curves below are
calculated  with the beam-size effect taken into account: the
bottom curve is actually two merged curves for sets 1 and 2 (the
set 1 is $\sigma_z=\Delta_z=50~\mu m, \sigma_y=\Delta_y=250~\mu
m, z_0=y_0=0$, set 2 is $\sigma_z=\Delta_z=50~\mu m,
\sigma_y=\Delta_y=250~\mu m, z_0=20~\mu m,y_0=0$); while the top
curve is for set 3 ($\sigma_z=\Delta_z=54~\mu m,
\sigma_y=\Delta_y=250~\mu m, z_0=y_0=0$). The data taken from
Fig.5c in \cite{P}.

\item {\bf Fig.5} The spectral intensity probability $\omega dw_{\gamma}/d\omega$
normalized to one particle in the beam in units $2\alpha
r_0^2\Sigma_z\Sigma_y/\pi$ versus the vertical displacement of
beams $z_0$ (in $nm$).

\end{itemize}

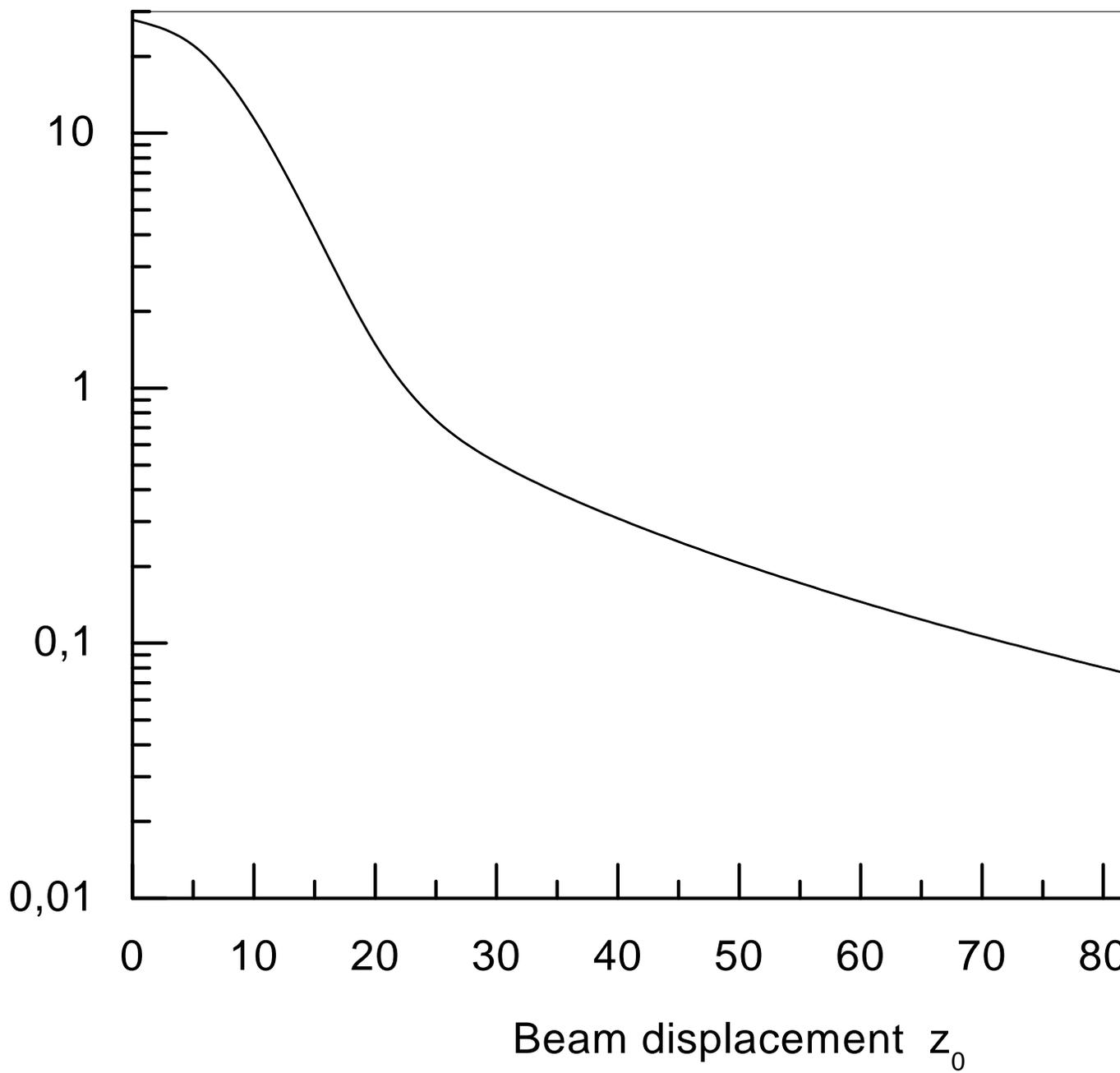
\end{document}